\documentstyle[11pt,newpasp,twoside,epsfig]{article}
\def\chem#1#2{$\rm{}^{#2}\kern-0.8pt#1$}
\def\reac#1#2#3#4#5#6{$\rm\,{}^{#2}\kern-0.8pt{#1}\,({#3}\,,{#4})\,
{}^{#6}\kern-0.8pt{#5}\,$}
\def\gsimeq{\,\,\raise0.14em\hbox{$>$}\kern-0.76em\lower0.28em\hbox  
{$\sim$}\,\,}  
\def\lsimeq{\,\,\raise0.14em\hbox{$<$}\kern-0.76em\lower0.28em\hbox  
{$\sim$}\,\,}  
\def\beqy{\begin{eqnarray}}
\def\eeqy{\end{eqnarray}}
\def\bmlet{\begin{mathletters}}
\def\emlet{\end{mathletters}}

\def\edcomment#1{\iffalse\marginpar{\raggedright\sl#1\/}\else\relax\fi}
\reversemarginpar

\begin{document}
\title{The nuclear processes responsible for the CNO synthesis}
 \author{M. Arnould, L. Siess and S. Goriely}
\affil{Institut d'Astronomie et d'Astrophysique, ULB - CP226, 1050 Brussels,
Belgium}
 
\begin{abstract}
The abundances of the isotopes of the elements C, N and O are mainly
affected by the cold CNO cycles in non-explosive stellar situations, or by
the hot CNO chains that can develop in certain explosive sites, like
classical novae. Helium burning phases can modify the composition of the
ashes of the CNO transmutations through several $\alpha$-capture reactions,
the most famed one being \reac{C}{12}{\alpha}{\gamma}{O}{16}. This
contribution presents a short review of the purely nuclear physics
limitations imposed on the accuracy of the predicted C, N and O yields from
H-burning in non-explosive stars or novae. This analysis makes largely use
of the NACRE compilation for the rates of the reactions on stable targets
making up the cold CNO cycle. Some more recent rate determinations are also
considered. The analysis of the impact of the rate uncertainties on the
abundance predictions is conducted in the framework of a simple parametric
astrophysical model. These calculations have the virtue of being a guide in
the selection of the nuclear uncertainties that have to be duly analyzed in
detailed model stars, particularly in order to perform meaningful
confrontations between abundance observations and predictions. They are
also hoped to help nuclear astrophysicists pinpointing the rate
uncertainties that have to be reduced most urgently. A limited use of
detailed stellar models is also made for the purpose of some specific
illustrations.
\end{abstract}

\section{Introduction}

As it is well known, the isotopes of the C, N and O elements derive their abundances at the stellar
surfaces and in the interstellar medium from the hydrogen and helium burning episodes taking place in
the central regions or in peripheral layers of all stars. These photospheric signatures may be
inherited from the stars at their birth, or result from so-called `dredge-up' phases, which are
expected to transport the H- or He-burning ashes from the deep production zones to the more external
layers. This type of surface contamination is encountered especially in low- and intermediate mass
stars on their first or asymptotic branches, where two to three dredge-up episodes have been
identified by stellar evolution calculations. Nuclear burning ashes may also find their
way to the surface of non-exploding stars by rotationally-induced mixing, which has been started to
be investigated in some detail, in particular by Andr\'e Maeder and his collaborators (Maeder \&
Meynet 2000), or by steady stellar winds, which have their most spectacular effects in massive stars
of the Wolf-Rayet type (these stars are very much beloved by Andr\'e; e.g. Maeder \&
Conti 1994). Non-explosive mass losses by Asymptotic Giant Branch or massive stars and catastrophic
supernova ejecta are especially efficient agents for the C, N and O budget of galaxies. 
 
The confrontation between calculated abundances and the wealth of observed elemental or isotopic C,
N and O abundances in stellar photospheres and in the interstellar medium can provide essential
clues to the stellar structure and evolution models, as well as to models for the chemical evolution
of the galaxies.  Of course, the information one can extract from such a confrontation is most
astrophysically useful if the discussion is freed from nuclear physics uncertainties to the largest
possible extent. 

Thanks to the impressive skill and dedication of some nuclear physicists, remarkable progress has
been made over the years in our knowledge of reaction rates at energies which are as close as
possible to those of  astrophysical relevance (e.g. Rolfs \& Rodney 1988). Despite these efforts,
important uncertainties remain. This relates directly to the enormous problems the experiments have
to face in this field, especially because the energies of astrophysical interest for
charged-particle-induced reactions are lower than the Coulomb barrier energies, especially in
non-explosive conditions. As a consequence, the corresponding cross sections can dive into the
nanobarn to picobarn abyss, which imposes to explore the `world of almost no event'. In general, it
has not been possible yet to measure directly such small cross sections. Theoreticians are thus
requested to supply reliable extrapolations from the lowest energies attained experimentally to
those of most direct astrophysical relevance. The problem is of a different nature in explosive
situations. The typical temperatures being higher in these cases, the energies of astrophysical
relevance come closer to the Coulomb barrier, which implies larger cross sections. However, there is
a high price to pay to enter this regime. The nuclear flows associated with stellar explosions
are indeed mostly located away from the valley of nuclear stability, which imposes the study of
reactions involving more or less highly unstable nuclei. The laboratory study of this `world of
exoticism' is another major challenge in nuclear astrophysics. 

Recently, a consortium of European laboratories has undertaken the difficult, but necessary, task of
setting up well documented and evaluated sets of experimental data or theoretical
predictions for a large number of astrophysically interesting nuclear reactions (Angulo et al.
1999). This compilation of reaction rates, referred to as NACRE (Nuclear Astrophysics Compilation
of REaction rates), comprises in particular the rates for all the
charged-particle-induced nuclear reactions involved in the non-explosive (`cold')  pp-, CNO, NeNa
and MgAl chains, the first two burning modes being essential energy producers, all four being
relevant nucleosynthesis agents. It also includes the most important reactions involved in
non-explosive helium burning as well as many other nuclear data of astrophysics interest, but of no
direct relevance here.

It has to be emphasized that the NACRE collaboration recommends the use in stellar evolution codes
of numerical reaction rates in tabular form. This philosophy differs markedly from the one promoted
by the previous widely used compilations (Caughlan \& Fowler 1988; hereafter CF88), and is expected
to lead to more accurate rate evaluation. However, as some stellar model builders still stick to the use of
analytical approximations, NACRE provides such formulae for the recommended rates. They differ in
several respects from the classically used expressions (e.g. CF88), the reasons for these changes
being discussed in Angulo et al. (1999).

The NACRE data are used in Sect.~2 to derive the abundances of the isotopes of C, N and O
involved in the cold CNO cycles. The abundance calculations are performed in the
framework of a simple parametric model that enlightens the impact of the rate
uncertainties on the derived abundances. These calculations have the virtue of being
free from blurring effects generated by the many intricacies of the stellar evolution
models. They thus help identifying in a clearer way purely nuclear uncertainties which
may have a significant impact on abundances predicted by detailed stellar models, or on
the build-up of models for the chemical evolution of galaxies. They are also hoped to
help nuclear astrophysicists pinpointing the rate uncertainties that have to be reduced
most urgently. Some post-NACRE data concerning \reac{N}{14}{p}{\gamma}{N}{15} and the
famed
\reac{C}{12}{\alpha}{\gamma}{O}{16} are briefly discussed in Sects.~3 and 4. Section~5
is devoted to an identification of the reactions of the explosive (`hot')
CNO cycles that may have an impact on C, N and O novae yields.
Brief conclusions are drawn in Sect.~6.

\begin{figure}
\vspace{-2.6cm}

\plotfiddle{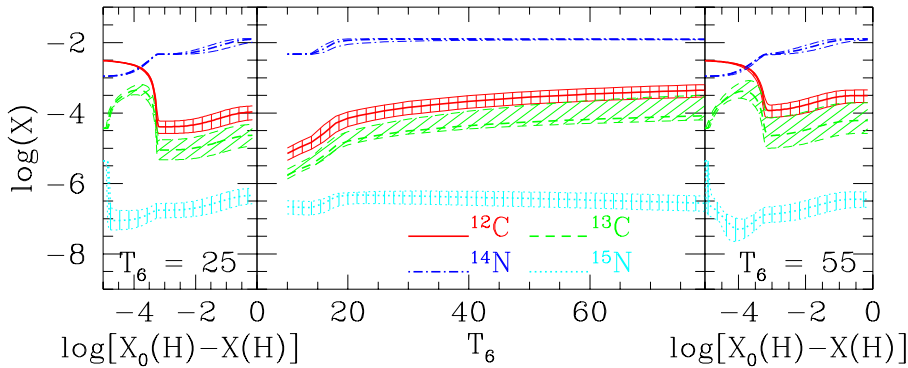}{8cm}{-90}{45}{45}{-175}{190}
\vspace{-0.8cm}
   \caption{ 
          {\it Left and right panels:} Time variations of the mass fractions of
          the stable C and N isotopes versus the amount of hydrogen burned at
          constant density $\rho=100\;{\rm g/cm^3}$ and constant 
          temperatures $T_6=25$ and 55. The H mass fraction is noted $X$(H), the
          subscript 0 corresponding to its initial value; 
          {\it Middle panel:} Mass fractions of the same nuclides at H exhaustion
          [X(H)=$10^{-5}$] as a function of $T_6$ [$T_6 \equiv T/10^6$]. The shaded areas
          delineate the uncertainties resulting from the reaction rates
          }
\vspace{-0.2cm}
\end{figure}

\section{The Cold CNO Cycles}

The reactions involved in the cold CNO cycles can be found in many places (e.g. Arnould, Goriely, \&
Jorissen 1999; hereafter AGJ), and need not be repeated here. It is well known that this H-burning
mode results in the production of \chem{He}{4} from H, and in the transformation  of the C, N and O
isotopes mostly into \chem{N}{14} as a result of the relative slowness of
\reac{N}{14}{p}{\gamma}{O}{15} with  respect to the other involved reactions. This
\chem{N}{14} build-up is clearly seen in Fig.~1.

The cold CNO cycles comprise branching points at \chem{N}{15}, \chem{O}{17}, and
\chem{O}{18}. In terms of the NACRE rates, their main characteristics may be very
briefly summarized as follows: (i) At $T_6=25$ ($T_6 \equiv T/10^6$),
\reac{N}{15}{p}{\alpha}{C}{12} is 1000 times faster than
\reac{N}{15}{p}{\gamma}{O}{16}, and the CN cycle reaches equilibrium already before $10^{-3}$ of the
initial protons have been burned; (ii) \reac{O}{17}{p}{\alpha}{N}{14} and
\reac{O}{17}{p}{\gamma}{F}{18} are the competing \chem{O}{17} destruction reactions.
The uncertainties in their rates have been strongly reduced in the last years, but remain quite
substantial. The rate of
\reac{O}{17}{p}{\alpha}{N}{14} recommended by NACRE is
larger than the CF88 one by factors of 13 and 90 at $T_6=20$ and 80, respectively. 
Smaller deviations, though reaching a factor of 9 at $T_6=50$, are found for the 
\reac{O}{17}{p}{\gamma}{F}{18} rate; (iii) the reactions \reac{O}{18}{p}{\gamma}{F}{19} and
\reac{O}{18}{p}{\alpha}{N}{15} compete at destroying \chem{O}{18}. At the temperatures of relevance, NACRE
predicts that \reac{O}{18}{p}{\gamma}{F}{19} is roughly 1000 times slower than
\reac{O}{18}{p}{\alpha}{N}{15}. However, at low temperatures, large uncertainties still affect the
\reac{O}{18}{p}{\gamma}{F}{19} rate.

The C, N and O isotopic compositions displayed in Figs.~1 and 2 are
calculated by assuming that H burning takes place at a constant density
$\rho = 100$ g\,cm$^{-3}$ and at two different constant temperatures, $T_6
= 25$ and 55.  The curves are constructed by combining in all possible ways
the lower and upper limits of all the relevant reaction rates. One
`reference' abundance calculation is also performed with all the
recommended NACRE rates. All the initial abundances are assumed to be solar
(Anders \& Grevesse 1989).  In spite of its highly simplistic aspect, this
analysis provides results that are of reasonable qualitative value, as
testified by their confrontation with detailed stellar model predictions.

Let us restrict ourselves here to some comments on the derived O isotopic composition and on the
production of \chem{F}{19}, which has long remained a puzzle in the theory of nucleosynthesis. As it
is well known, the O isotopic composition depends drastically on the burning temperature.  In
particular, \chem{O}{17} is produced at $T_6 \lsimeq 25$, but is destroyed at higher temperatures.
This has the important consequence that the amount of \chem{O}{17} emerging from the CNO cycles and
eventually transported to the  stellar surface is a steep function of the stellar mass. This
conclusion could get some support from the observation of a large spread in the oxygen isotopic 
ratios at the surface of red giant stars of somewhat different masses (Dearborn 1992, and references
therein). Figure~2 also demonstrates that the oxygen isotopic composition cannot be fully reliably
predicted yet at a given temperature as a result of the cumulative uncertainties associated with the
different production and destruction rates. This situation prevents any meaningful comparison to be
made with spectroscopic data, and, even more so, any firm conclusion to be drawn from models for the
chemical evolution of galaxies. 

\begin{figure}[t]
\vspace{-1.3cm}
\plotfiddle{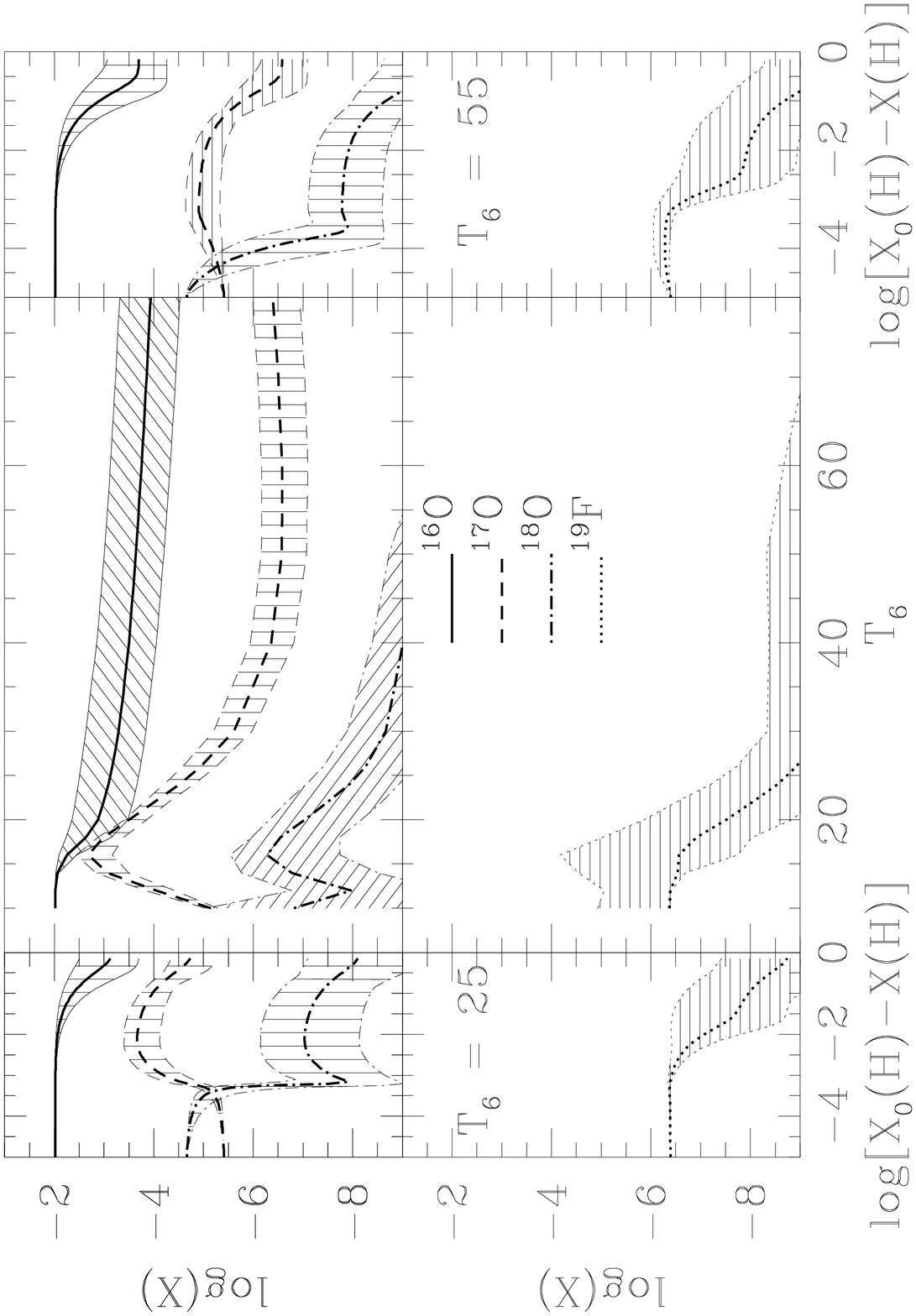}{8cm}{-90}{32}{32}{-130}{220}
\vspace{-1.0cm}
\caption{Same as Fig.~1, but for the O and F nuclides}
\end{figure}

As far as \chem{F}{19} is concerned, only its solar system abundance has been known
for a long time. The observational situation has changed substantially with the measurements by
Jorissen, Smith, \& Lambert (1992) of fluorine at the surface of red giant stars considered to be in
their post-first dredge-up phase, as well as of AGB stars, some of the analyzed stars exhibiting a F
enhancement with respect to solar. Theoretically, the slowness of the recommended NACRE rate for
\reac{O}{18}{p}{\gamma}{F}{19} relative to the one of the competing
\reac{O}{18}{p}{\alpha}{N}{15} reaction undermines the path leading to the production of
\chem{F}{19} from oxygen. However, the NACRE upper bound for the proton radiative capture by
\chem{O}{18} could be comparable to the \reac{O}{18}{p}{\alpha}{N}{15} rate, and at the same time
larger than the \reac{F}{19}{p}{\alpha}{O}{16} rate at $T_6 \lsimeq 20$. As a result, some 
\chem{F}{19} might be produced in the wake of the CNO cycles, in contradiction with the conclusion 
drawn from the adoption of the CF88 rates. Figure~2 indeed confirms that fluorine  could be
overproduced (with respect to solar) by up to a factor of 100 at H exhaustion when $T_6 \approx
15$.  However, Fig.~2 also reveals that the maximum \chem{F}{19} yields that can be attained remain
very poorly predictable as a result of the rate uncertainties. 
In fact, some hint of a non-negligible production of fluorine by the CNO cycles might come
from the observation of fluorine abundances slightly larger than solar at
the  surface of red giant stars considered to be in their post-first dredge-up  phase
(Jorissen et al. 1992; Mowlavi, Jorissen, \& Arnould 1996). Other possibilities of
significant production of \chem{F}{19} have also been identified, the most promising ones
being shell He burning in AGB stars [and more specifically the partial
mixing of protons in the C-rich layers at the time of the third dredge-up (Goriely \&
Mowlavi 2000) which could account for the F overabundance in AGB
stars observed by Jorissen et al. 1992] or central He burning in Wolf-Rayet stars
(Meynet
\& Arnould 1996, 1999; Mowlavi, Jorissen,
\& Arnould 1998). 
Finally, let us note that any important leakage out of the CNO cycles to \chem{Ne}{20} is prevented
by the fact that \reac{F}{19}{p}{\alpha}{O}{16} is always much faster than
\reac{F}{19}{p}{\gamma}{Ne}{20}, this conclusion being independent of the remaining rate
uncertainties.

\section{Post-NACRE Data for \reac{N}{14}{p}{\gamma}{O}{15}: No Revolution in Stellar
Evolution}

The  \reac{N}{14}{p}{\gamma}{O}{15} reaction is the slowest of the CN
cycle. It is thus expected to play a special role in the structure,
evolution and concomitant nucleosynthesis of stars in which the burning
mode is not dominated by the p-p chains (i.e. $M \ga
1.3$M$_\odot$ in case of solar metallicity).

The estimate of the \reac{N}{14}{p}{\gamma}{O}{15} reaction cross section
at low energies has recently been revisited by Angulo et al. (2001). As in
the NACRE compilation, the experimental data of Schr\"oder et al. (1987)
are adopted for the calculation of the non-resonant contribution to the
reaction rate. The rate is estimated to be about 70\% smaller than the
NACRE one at temperatures $T_6 \la 150$, and about twice larger above
(Fig.~3). This significant difference originates from the new $R$-matrix
calculation of the direct transition to the ground-state which is
responsible for the major contribution to the low-energy S-factor. The
\reac{N}{14}{p}{\gamma}{O}{15} reaction has also been re-analyzed
experimentally using the inverse kinematics approach (Galloy \& Terwagne,
private communication).

\begin{figure}
\vspace{-2.4cm}
\plotfiddle{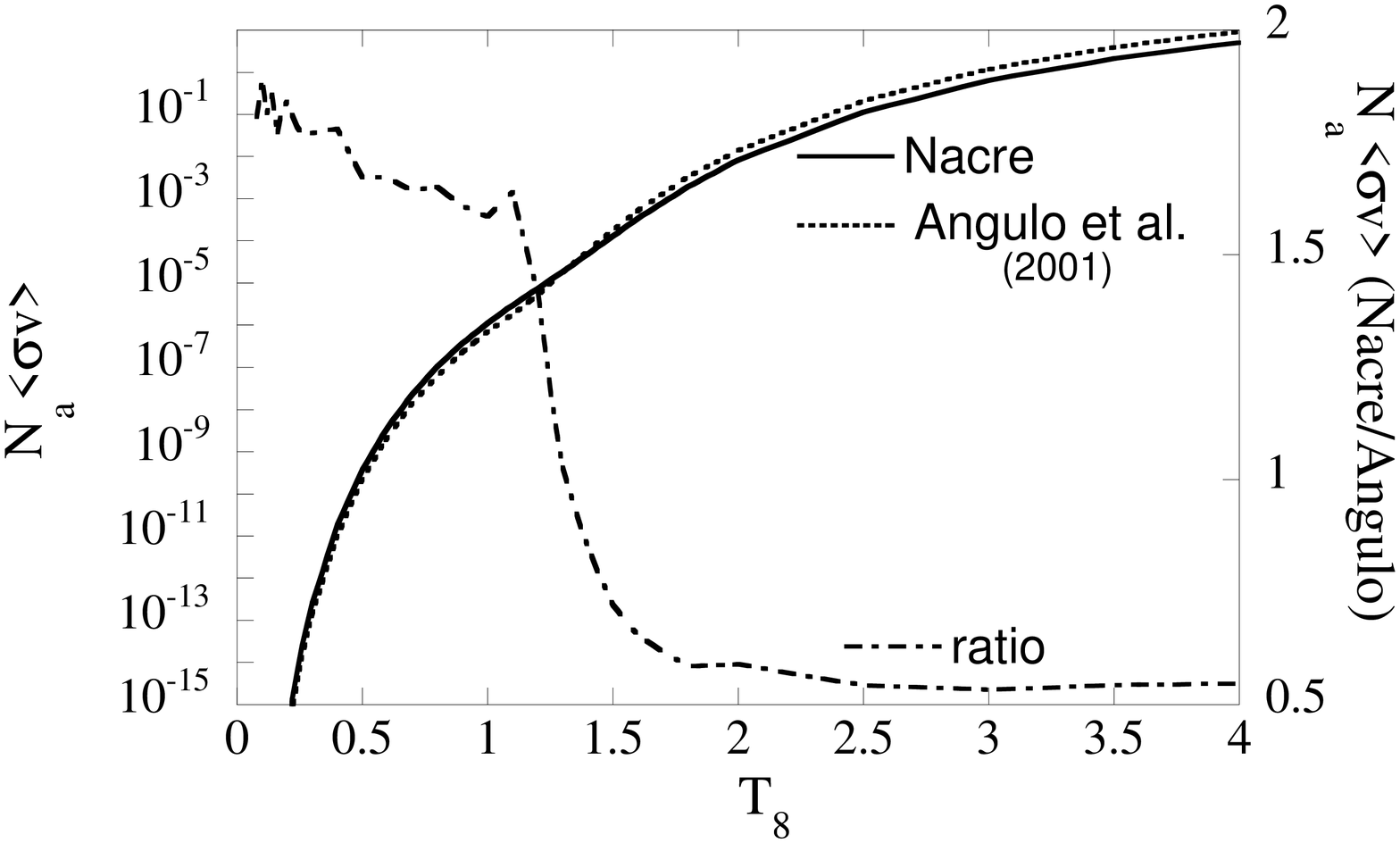}{8cm}{0}{35}{35}{-150}{0}
\vspace{-0.8cm}
\caption[]{Comparison of the reaction rate for $^{14}$N(p,$\gamma$)$^{15}$O
between the adopted NACRE rate and the new determination of Angulo et
al. (2001) as a function of the temperature [$T_8 \equiv T/(10^8)$~K]. The
dot-dash line gives the ratio of the two rates, to be read on the right
axis.}
\end{figure}
 
In spite of the bottleneck role of \reac{N}{14}{p}{\gamma}{O}{15}, its rate
changes displayed in Fig.~3 have only a very small impact on the 5 and 20
M$_\odot$ star models we have tested. The use of the faster rate instead of
the slower 2001 rate increases the duration of the main sequence by only
$\sim 2$\% as a result of a smaller central temperature. This difference 
cannot be identified observationally.  The effects on the central CNO
abundances of these two different rates are illustrated in Fig.~4. Obviously,
using the highest (NACRE) rate favors the production of all the species
entering the CN cycle at the expense of \chem{N}{14}. However, these
changes barely affect the surface composition after the first dredge
up. The most prominent difference is a 5\% increase in the
\chem{N}{15}/\chem{N}{14} ratio when using the fastest rate. As a
conclusion, no decisive observational effects result from the use the two
rates of Fig.~3 and the effects on the evolution are extremely weak.
\begin{figure}
\vspace{-0.6cm}
\plotfiddle{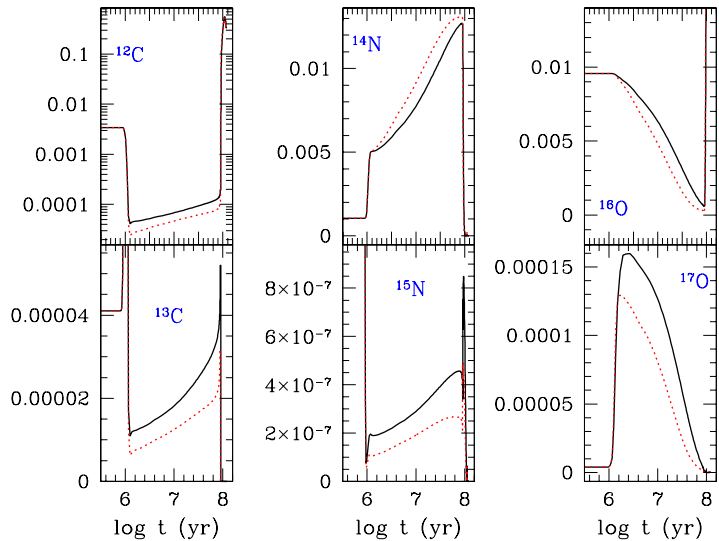}{10.5cm}{0}{50}{50}{-160}{-10}
\vspace{-3.2cm}
\caption[]{Evolution of the central mass fraction of the main CNO nuclei in
a 5 M$_\odot$ stellar models of solar metallicity. The solid and dotted
lines refer to calculations performed using the Nacre and Angulo et
al. reaction rates for $^{14}$N(p,$\gamma$)$^{15}$O reaction,
respectively}
\vspace{-0.2cm}
\end{figure}

\section{The \reac{C}{12}{\alpha}{\gamma}{O}{16}: Where do we stand?}

This reaction has long been recognized as essential in many stellar physics questions. The precise
knowledge of its rate is indeed a {\it necessary} condition for a reliable evaluation of the
\chem{C}{12} to \chem{O}{16} abundance ratio at the time He is exhausted by non-explosive burning,
and consequently of the progeny of the subsequent non-explosive or explosive (supernova) C or
O burnings. Even the precise fate of massive stars at the end of their evolution is influenced by the
rate of the transformation of \chem{C}{12} into \chem{O}{16}, which, needless to say, is also not
beside the point of describing the chemical evolution of galaxies.  Of course, the knowledge of this
rate is in no way {\it sufficient} to provide high quality predictions for all the above matters,
stellar or galactic models having many difficulties of their own. The reader is referred to e.g.
Imbriani et al. (2001) for a discussion of the impact of the \reac{C}{12}{\alpha}{\gamma}{O}{16}
rate on the modeling of stars in a wide mass range.

The nuclear physics data concerning \reac{C}{12}{\alpha}{\gamma}{O}{16} are summarized by e.g.
Angulo et al. (1999). Since the NACRE compilation, many laboratory efforts have been devoted to
reducing its rate uncertainties, but the challenge is impressive. No measurement can be foreseen
at the energies of astrophysical relevance, and the extrapolation to them from the lowest energy
measurements is especially risky. The reaction mechanism presents a
remarkable accumulation of intricacies, the existence of two sub-threshold states and their
interferences with states above threshold being clearly not the least.  Kunz et al. (2002) have
recently published new direct measurements, from which Fig.~5 is extracted. At the
non-explosive He-burning temperatures ($T$ around $10^8$ K), their recommended
rate is not substantially different from the NACRE adopted values. However,
uncertainties remain. 

Some had the dream of constraining the \reac{C}{12}{\alpha}{\gamma}{O}{16} rate by confronting the
solar system composition with the predictions from models for the chemical evolution of the Galaxy. A
new fashion has also developed recently, based on the precision study of white dwarf pulsations. From
these, the hope has been expressed to derive the interior C/O abundance ratios accurately, and so to
get accurate \chem{C}{12} transmutation rates. We consider that these astrophysics methods to obtain
a reliable information on the microphysics of the $\alpha$-capture rate represents an impossible
inverse problem. We advise to better go to the nuclear physics laboratory, even if the task is far
from being trivial!

\begin{figure*}
\vspace{-0.6cm}
\plotfiddle{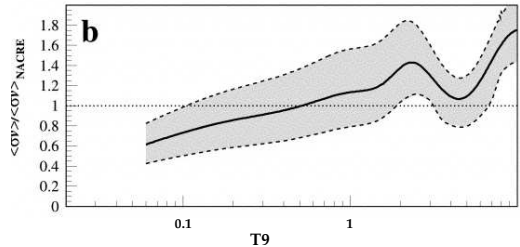}{9cm}{0}{35}{35}{-120}{155}
\vspace{-5.6cm}
\caption{Comparison between the \reac{C}{12}{\alpha}{\gamma}{O}{16} stellar rates
derived by Kunz et al. (2002) and the
adopted NACRE rates versus temperature [$T_9 \equiv T/(10^9)$K]. The shaded
area corresponds to the rate uncertainties assigned by Kunz et al. (2002)}
\end{figure*}

\section{Novae and the Hot CNO Burning}

Hydrogen can burn explosively in various astrophysical events, like novae or x-ray bursts. The
corresponding hot burning modes (hot p-p and CNO or NeNa-Mg-Al chains, rp- or $\alpha$p processes)
involve a variety of unstable nuclei, and have specific nucleosynthesis signatures. They raise many
difficult experimental and theoretical nuclear physics questions, and highly intricate astrophysical
problems as well. They develop when some produced $\beta$-unstable nuclei decay more slowly than they
capture protons, which is in contrast to the situation characterizing the corresponding cold burnings.

Of concern here is the hot CNO chain, first introduced by Audouze, Truran, \& Zimmerman (1973) and
Arnould \& Beelen (1974). The cold CNO cycles switch to the hot mode when
\reac{N}{13}{p}{\gamma}{O}{14} becomes faster than the \chem{N}{13} $\beta$-decay. This occurs
typically at temperatures in excess of $10^8$ K, which can be attained in thermonuclear runaways of
the classical nova type. Such  explosions of such a type are predicted to occur as a result of
the accretion at a suitable rate onto a C-O or O-Ne white dwarf of material from a companion in a
binary system. A sensitivity analysis of the nova yields to reaction rate uncertainties has been
conducted recently by Iliadis et al. (2002). The uncertainties in the rates of
\reac{O}{17}{p}{\gamma}{F}{18}, \reac{O}{17}{p}{\alpha}{N}{14} and \reac{F}{18}{p}{\alpha}{O}{15} are
found to be responsible for variations by at least a factor of 2 in the C-O nova yields of
\chem{O}{17} and \chem{F}{18}. 
Let us just note that the proton capture rates on \chem{O}{17}
have been re-examined recently by Blackmon (private communication) for typical nova conditions. They
are also uncertain in cold CNO burning regimes (Sect.~2). On the other hand, the radionuclide
\chem{F}{18} ($t_{1/2} = 110$ min) predicted to be produced in novae is considered by some as
important for $\gamma$-ray astrophysics. The annihilation with electrons of the positrons emitted in
its $\beta^+$-decay and Compton scattering produce a $\gamma$-ray radiation that
might be observable when the expanding nova envelope becomes transparent to
$\gamma$-rays in the relevant energy range (e.g. Hernanz et al. 1999).
 
\section{Conclusions}

As an aid to the confrontation between spectroscopic observations and
theoretical expectations, the nucleosynthesis associated with the cold CNO
cycles is studied with the help of the recent NACRE compilation of nuclear
reaction rates. Special attention is paid to the impact on the derived
abundances of the carefully evaluated uncertainties that still affect the
rates of many reactions. In order to isolate this nuclear effect in an
unambiguous way, a very simple constant temperature and density model is
adopted. Some post-NACRE data are also discussed briefly, sometimes in the
framework of detailed model stars. Finally, a recent analysis of the impact
of the uncertainties in the rates of the hot CNO chain on the C, N and O
yields from classical novae is summarized.

It is shown that large spreads in the abundance predictions for several nuclides may result
not only from a change in temperature, but also from nuclear physics uncertainties. This
additional intricacy has to be kept in mind when trying to interpret the observations and
when attempting to derive constraints on stellar models from these data.

\acknowledgements
{S.G. is F.N.R.S research associate. L.S acknowledges support from a TMR
  ``Marie-Curie Fellowship''at ULB}

\end{document}